\documentclass[twocolumn, pra]{revtex4-1}

\usepackage{natbib}
\usepackage{graphicx}
\usepackage{dcolumn} 
\usepackage{bm}
\usepackage{hyperref}
\usepackage{enumitem}
\usepackage[mathlines]{lineno}
\usepackage[titletoc, title]{appendix}

\begin{document}

\title{Two-species boson mixture on a ring: a group theoretic approach to the quantum dynamics of low-energy excitations}

\author{Vittorio Penna}
\author{Andrea Richaud}
\email{andrea.richaud@polito.it}
\affiliation{
Dipartimento di Scienza Applicata e Tecnologia and u.d.r. CNISM, Politecnico di Torino, 
Corso Duca degli Abruzzi 24, I-10129 Torino, Italy}
\date{\today}

\begin{abstract}
We investigate the weak excitations of a system made up of two condensates 
trapped in a Bose-Hubbard ring and coupled by an interspecies repulsive interaction. 
Our approach, based on the Bogoliubov approximation scheme, shows that one can reduce the problem Hamiltonian to the sum of sub-Hamiltonians $\hat{H}_k$, each one associated to momentum modes $\pm k$.
Each $\hat{H}_k$ is then recognized to be an element of a dynamical algebra. This uncommon and remarkable property allows us to present a straightforward diagonalization scheme, to find constants of motion,
to highlight the significant microscopic processes, and to compute their time evolution. 
The proposed solution scheme is applied to a simple but non trivial closed circuit, the trimer. 
The dynamics of low-energy excitations, corresponding to weakly-populated vortices, is investigated
considering different choices of the initial conditions, and the angular-momentum transfer between the 
two condensates is evidenced. Finally, the condition for which the spectral collapse and 
dynamical instability are observed is derived analytically.
\end{abstract}

\maketitle

\section{Introduction}
\label{sec:Introduction}

Bosonic mixtures formed by two atomic species have proved to be a fertile ground for investigating the complex interplay of the density-density interspecies interaction with the interactions among bosons of the same species, boson hopping, and interspecies population imbalance.
Trapped in optical lattices where the space 
fragmentation emphasizes the quantum nature of microscopic processes, mixtures 
have revealed an extraordinarily rich scenario of new exotic quantum phases, dynamical behaviors and properties. 
These include, for example, unprecedented Mott-like states and types of
superfluidity \cite{mix1}-\cite{mix3}, glassy phases \cite{mix4},  interspecies coherence \cite{mix5}, spatial separation or coexistence of different phases \cite{mix6}, \cite{mix7}, and polaron-like excitations \cite{mix8}.
Dynamical properties have also been explored within the mean-field picture and a variety of aspects has been considered which range from the dynamical stability of binary mixtures \cite{dmix0}, different types of self-trapping solutions \cite{dmix1}, and the effectiveness of the space-mode description in the Gross-Pitaevskii picture \cite{dmix2}, to the dynamics of the Rabi-Josephson regime \cite{dmix3}, and quantum many-body correlations \cite{dmix4}.
More recent work has been focused on the interspecies coherent transfer of superfluid vortices \cite{dyn1}, the analytic description of spectral collapse \cite{dyn2} and the miscibility of a two-component mixture \cite{dyn3}. 
Mixtures, well described within the Bose-Hubbard (BH) picture \cite{mix9}, \cite{mix10}, have been realized by means of either two atomic species \cite{mix11} or the same species in two different internal states \cite{mix12}.

While dynamical aspects of mixtures have been mainly studied through the mean-field approach and in the simple geometry of a double-well potential \cite{dmix0}-\cite{dyn3}, minor attention has been devoted to extended arrays of wells and the use of techniques better incorporating the deep quantum character of such systems. In this article, using the method applied to the dynamics of bosons in a two-ring ladder \cite{Noi}, we analyze the low-energy excitations in a two-component bosonic mixture described within the BH picture and confined on ring lattice. The ring geometry, designed more than ten years ago in \cite{am1ring}, has recently raised a lot of interest in the study of atomtronic devices \cite{am2ring}, \cite{am3ring} and its feasibility has been proved to be within the reach of current experimental techniques.

We focus on the regime where both condensed species are superfluid and uniformly distributed in the ring lattice. This allows us to reduce the system Hamiltonian to an effective quantum model derived by means of the Bogoliubov approximation, which well describes the low-energy excitations in the uniform-density regime. 
After observing that the original Hamiltonian decouples into many sub-Hamiltonians (involving \textit{pairs} of opposite-momentum modes), we apply a well-established group-theoretic procedure based on the identification of the dynamical algebra of a system \cite{MR}, \cite{ZFG}. In fact, we recognize that the dynamical algebra of each sub-Hamiltonian is so(2,3). This important property provides a viable, fully-analytic diagonalization scheme, allows one to find conserved quantities and highlights the most important microscopic processes that underlie the dynamical evolution of the system. 

We apply this quite general solution scheme to a three-site BH model (trimer) with a ring geometry to find the energy spectrum of low-energy excitations as a function of the model parameters and the Heisenberg equations governing the dynamics of the system.
Then, we study the trimer-excitation dynamics for different values of Hamiltonian parameters  and initial conditions relevant to simple but significant rotational states. Interestingly, we observe that angular momentum can indeed be transferred between the two condensed species and we point out the condition under which this phenomenon occurs. 
Finally, we focus our attention on the stability of the system and derive the analytic parameter-dependent formula giving the critical condition at which the energy-spectrum collapse takes place and the system turns unstable. This circumstance can reveal the presence of a phase transition, an important point that will be studied in a future article. 

Our work is organized as follows: in Section \ref{sec:Model_presentation} we introduce the model and, by implementing the Bogoliubov approximation, we reduce it to a form charaterized by a well-defined dynamical algebra. The derivation of the excitations spectrum and of the Heisenberg equations is described in Section \ref{sec:Dynamical_algebra}. In Section \ref{sec:Mixture_on_a_trimer} we apply the dynamical-algebra method to the trimer and investigate the dynamics of boson populations in various regimes. Section \ref{sec:Stability} includes the stability analysis and, finally, Section \ref{sec:Concluding_remarks} is devoted to concluding remarks.

\section{Ring-mixture Hamiltonian and dynamical algebra approach}
\label{sec:Model_presentation}
The Hamiltonian describing a mixture
of two condensates loaded in a ring lattice and coupled by an interspecies 
repulsive term is described, in the formalism of second-quantization, by
$$
  \hat{H}= \frac{U_a}{2} \sum_{j=1}^{L} N_j(N_j-1)- T_a \sum_{j=1}^{L} \left(A_{j+1}^\dagger A_j +A_j^\dagger A_{j+1} \right) 
$$
$$
  +\frac{U_b}{2} \sum_{j=1}^{L} M_j(M_j-1)- T_b \sum_{j=1}^{L} \left(B_{j+1}^\dagger B_j +B_j^\dagger B_{j+1} \right)
$$
\begin{equation}
\label{eq:interaction}
+W \sum_{j=1}^{L} N_j\, M_j ,
\end{equation}
where $T_a$ and $T_b$ represent the hopping amplitudes, $U_a$ and $U_b$
the \textit{intra}-species repulsive interactions, and $W$ is the
\textit{inter}-species repulsion. $A_j$ and $B_j$ are standard bosonic operators, featuring the commutators $[A_j,A_k^\dagger]=\delta_{j,k}=[B_j, B_k^\dagger]$, and $[A_j,B_k^\dagger]=0$. The particle number $N = \sum_j N_j$ and $M = \sum_j M_j$ is conserved in both condensates. $N_j=A_j^\dagger A_j$ and $M_j=B_j^\dagger B_j$ are number operators and $L$ is the site number in the ring lattice. 

By assuming $T_a$ ($T_b$) sufficiently larger than $U_a$ ($U_b$) in order to avoid the emergence of Mott-insulator states, the condition that $W$ is sufficiently smaller than $U_a$, $U_b$ implies that the mixture ground state is a superfluid in which the two components are completely mixed and delocalized (see, for example, \cite{mix6}) with local populations 
$N_j = N/L$ and $M_j = M/L$.
As is well known, this uniform boson distribution entails macroscopically-occupied
zero-momentum modes for both species within the momentum-mode picture and suggests
the application of the Bogoliubov approximation. 
In order to properly enact it, we take advantage of the ring-structure of the system, and move to momentum-mode basis. Momentum-mode operators, $a_k$ and $b_k$, are defined in terms of
site-modes operators
$$
 A_j=\sum_{k=1}^{L} \frac{a_k}{\sqrt{L}} e^{+i\tilde{k}aj}, \qquad B_j=\sum_{k=1}^{L} \frac{b_k}{\sqrt{L}} e^{+i\tilde{k}aj}, 
$$
with $\tilde{k}= ({2\pi}/{d}) k$ and $d = L a$. 
Parameter $a$ is the lattice constant, $d$ is 
the ring length, $N$ ($M$) is the number of atoms of atomic species A (B) and the summations are restricted to the first Brillouin zone. 
Momentum-mode operators $a_k$ and $b_k$ are also characterized by standard bosonic commutators
$[a_j,a_k^\dagger]=\delta_{j,k}$, $[b_j,b_k^\dagger]=\delta_{j,k}$ and $[a_j,b_k^\dagger]=0$.
In view of the ring geometry, the number of bosons having (angular) momentum equal to $\hbar \tilde k$ is associated to number operators $n_k=a_k^\dagger a_k$ and $m_k=b_k^\dagger b_k$. In the momentum-mode picture, 
the Hamiltonian can be recast into the form
$$
\hat{H}= \frac{U_a}{2L} \sum_{p,q,k=1}^{L}
a_{q+k}^\dagger a_{p-k}^\dagger a_q a_{p} -2T_a\sum_{k=1}^{L} a_k^\dagger a_k \; \cos(a\tilde{k}) 
$$
$$
 +\frac{U_b}{2L} \sum_{p,q,k=1}^{L}    b_{q+k}^\dagger b_{p-k}^\dagger b_q b_{p} -2T_b\sum_{k=1}^{L} b_k^\dagger b_k \, \cos(a\tilde{k}) 
$$  
$$
+ \frac{W}{ L}\sum_{p,q,k=1}^{L}  a_{p+k}^\dagger\, b_{q-k}^\dagger\, a_p\, b_q .
$$
If the system is in the superfluid phase ($T_\nu/U_\nu$, $\nu=a, b$ large enough) 
and both species are uniformly distributed in the lattice (small $W/U_\nu, \nu=a,b$ 
excludes spatial phase separation), both atomic species are characterized by a macroscopically occupied momentum mode, namely $r=0$, and the Bogoliubov procedure \cite{Rey, Burnett} can be easily enacted (details are given in the Supplemental Material).
The resulting Hamiltonian
\begin{equation}
\label{eq:Decoupling}
    \hat{H}= E_0 + \sum_{k>0} \hat{H}_k ,
\end{equation}
can be written as the sum of a constant term $E_0={u_a}(N-1)/2-2NT_a+ {u_b}(M-1)/2-2MT_b +w\sqrt{NM} - \sum_{k>0}(\gamma_{k,a}+\gamma_{k,b})$, and $(L-1)/2$ decoupled Hamiltonians 
$\hat{H}_k$, each one involving just one \textit{pair} of (opposite) momentum modes,
$$
   \hat{H}_k =  2 (\gamma_{k,a} A_3 + \gamma_{k,b} B_3)  
+u_a \left( A_+ +A_-\right) 
$$
\begin{equation}
\label{eq:H_k}
+u_b \left(B_++B_-\right)+ w (K_++K_-+S_++S_-).
\end{equation}
where 
$$
A_+ = a_{k}^\dagger a_{-k}^\dagger,
\qquad
B_+ = b_{k}^\dagger b_{-k}^\dagger,
$$
\begin{equation}
\label{eq:A3B3}
 A_3 = \frac{n_{k}+n_{-k}+1}{2}, \qquad B_3 = \frac{m_{k}+m_{-k}+1}{2},
\end{equation}
$$
 S_+ = a_{k}^\dagger b_{k}+a_{-k}^\dagger b_{-k}, \qquad  K_+ = a_{-k}^\dagger b_{k}^\dagger +a_{k}^\dagger b_{-k}^\dagger.
$$
Note that operators $X_-$ (with $X= A, B, S, K $) are simply given by $X_-=(X_+)^\dagger$, 
while we define 
$$
\gamma_{k,a}=u_a-2T_a \left ( c_k-1\right ), 
\quad 
\gamma_{k,b}=u_b-2T_b \left ( c_k-1\right ),
$$
$$
u_a=\frac{U_a N}{L}, \qquad u_b=\frac{U_b M}{L}, \qquad  w=\frac{W\sqrt{NM}}{L},
$$
with $c_k =\cos ( a\tilde{k} )$, to lighten the notation.

An essential feature of Hamiltonian (\ref{eq:H_k}) is that the set of two-mode operators $\mathcal{A}=\{A_\pm, \, B_\pm,\, K_\pm,\, S_\pm,\, A_3,\, B_3\}$ is closed under commutation. The recognition of this set, the so-called \textit{dynamical algebra} \cite{MR, ZFG}, is particularly advantageous for computing the conserved physical quantities, for a straightforward diagonalization of Hamiltonian $H_k$ and for deriving the Heisenberg equations of the system. As regards our problem, the dynamical algebra generated by $\mathcal{A}$ is so(2,3). Its characteristic commutators and details on the dynamical-algebra method can be found in \cite{Noi}.

Interestingly, the same algebra was found to feature the completely different model of a single condensate trapped in a two-ring ladder \cite{Noi}. In that model, terms $S_\pm$ described the angular-momentum transfer between two coupled rings via tunnelling effect. In the present model, in addition to terms $S_\pm$, the new operators $K_\pm$ appear which describe creation and destruction processes that were absent in the two-ring model. This difference comes from the inter-species coupling term of model (\ref{eq:interaction}) which replaces the inter-ring tunneling term of the two-ring model \cite{Noi}. 
 
Concluding, we note that this approach is valid also in the case $U_a,U_b<0$, describing 
attractive interactions among bosons of the same species, provided 
that {\it i)} the ratio $T_\nu/|U_\nu|$ ($\nu=a,b$) is sufficiently large, in order to guarantee a 
superfluid phase, and {\it ii)} the ratio $W/|U_\nu|$ is small enough to determine uniform boson distributions \cite{Jack} apt to apply the Bogoliubov scheme.

\section{Diagonalization of the model}
\label{sec:Dynamical_algebra}
In order to diagonalize our model, we choose the set $\{| n_k, \,n_{-k},\, m_k, \, m_{-k}\rangle\}$ as a basis of the Hilbert space of states associated to sub-Hamiltonian $\hat{H}_k$. The four quantum numbers $n_{\pm k}$ and $m_{\pm k}$ which label a basis vector, correspond to the numbers of bosons endowed with angular momentum $\pm k$ in either the condensed species. Since the numbers of particles in the two species are conserved, the number of bosons in momentum modes $r=0$ (modes that have been made semiclassical) are
$$
n_0=N-\sum_{k\neq 0} n_k , \qquad m_0=N-\sum_{k\neq 0} m_k.
$$    

\subsection{Energy eigenstates and eigenvalues}
\label{sub:Spectrum}
The information that a Hamiltonian belongs to a certain dynamical algebra allows one to calculate its spectrum in a straightforward way. In general, one can exploit the algebra structure to numerically diagonalize the Hamiltonian. In the particular but significant case where $u_a=u_b=:u$ and $T_a=T_b=:T$, the diagonalization process is fully analytic, and sub-Hamiltonian 
$$
  \hat{H}_k = 2\gamma_{k} \left(A_3 +  B_3\right)  +u \left( A_+ +A_- + B_++B_-\right )
$$
$$
 + w (K_++K_-+S_++S_-)
$$
can be put in diagonal form by making use of a unitary transformation $U_k$ belonging to group SO(2,3)
$$
U_k =  e^{\frac{\varphi}{2}(S_- -S_+)}
e^{\frac{\theta_a}{2}(A_- -A_+)} e^{\frac{\theta_b}{2}(B_- -B_+)}.
$$
A proper choice of angles $\varphi$, $\theta_a$ and $\theta_b$, allows one 
to get rid of those operators which are not diagonal in the Fock-states basis. By choosing
\begin{equation}
\label{eq:Rotation_angles}
\varphi=\frac{\pi}{2}, 
\qquad 
\mathrm{th} \; \theta_a =\frac{u+w}{\gamma_k +w}, 
\qquad 
\mathrm{th}\,\theta_b =\frac{u-w}{\gamma_k -w},
\end{equation}
one obtains the diagonal Hamiltonian
$$
\hat{\mathcal{H}}_k = 
U_k^{-1}\,\hat{H}_k\,U_k= 2A_3\sqrt{(\gamma_k-u)(\gamma_k+u+2w)}  
$$
\begin{equation}
\label{eq:Diagonal_Hamiltonian}
+2B_3\sqrt{(\gamma_k-u)(\gamma_k+u-2w)}
\end{equation}
in which operators $A_3$ and $B_3$ are linear combinations of number operators 
(see formulas (\ref{eq:A3B3})), and thus are diagonal in the Fock-states basis.
Since $\hat{H} =\sum_k \hat{H}_k$, then
the energy spectrum of $\hat{H}$ takes the form
$$
E(\{n_k,n_{-k},m_k,m_{-k}\}) = 
$$
$$
-4TN +\frac{U}{L}(N^2-N) +\frac{W}{L}N^2 
+\sum_{k>0}\biggl[4T(c_k-1)-2u  \biggr.
$$
\begin{equation}
\label{eq:spectrum}
\biggl. +\hbar \omega_k (n_k+n_{-k}+1  )+ \hbar \Omega_k (m_k+m_{-k}+1) \biggr]
\end{equation}
where $\tilde{k}= {2\pi k}/{(aL)}$ in $c_k=\cos(a\tilde{k})$, $k\in \left[1,({L-1})/{2}\right]$ is an integer index, and
\begin{equation}
\label{eq:omega_k}
\omega_k = \frac{1}{\hbar}\sqrt{2T(1-c_k)[2T(1-c_k)+2u+2w]},
\end{equation}
\begin{equation}
\label{eq:Omega_k}
\Omega_k = \frac{1}{\hbar}\sqrt{2T(1-c_k)[2T(1-c_k)+2u-2w]}.
\end{equation}
are characteristic frequencies associated to each Hamiltonian $\hat{H}_k$ together with quantum numbers $n_{\pm k}$ and $m_{\pm k}$.

\subsection{Time-evolution of physical observables and microscopic processes}
\label{sub:Linear_system}
\begin{figure}[h!]
\includegraphics[angle=270,width=1\columnwidth]{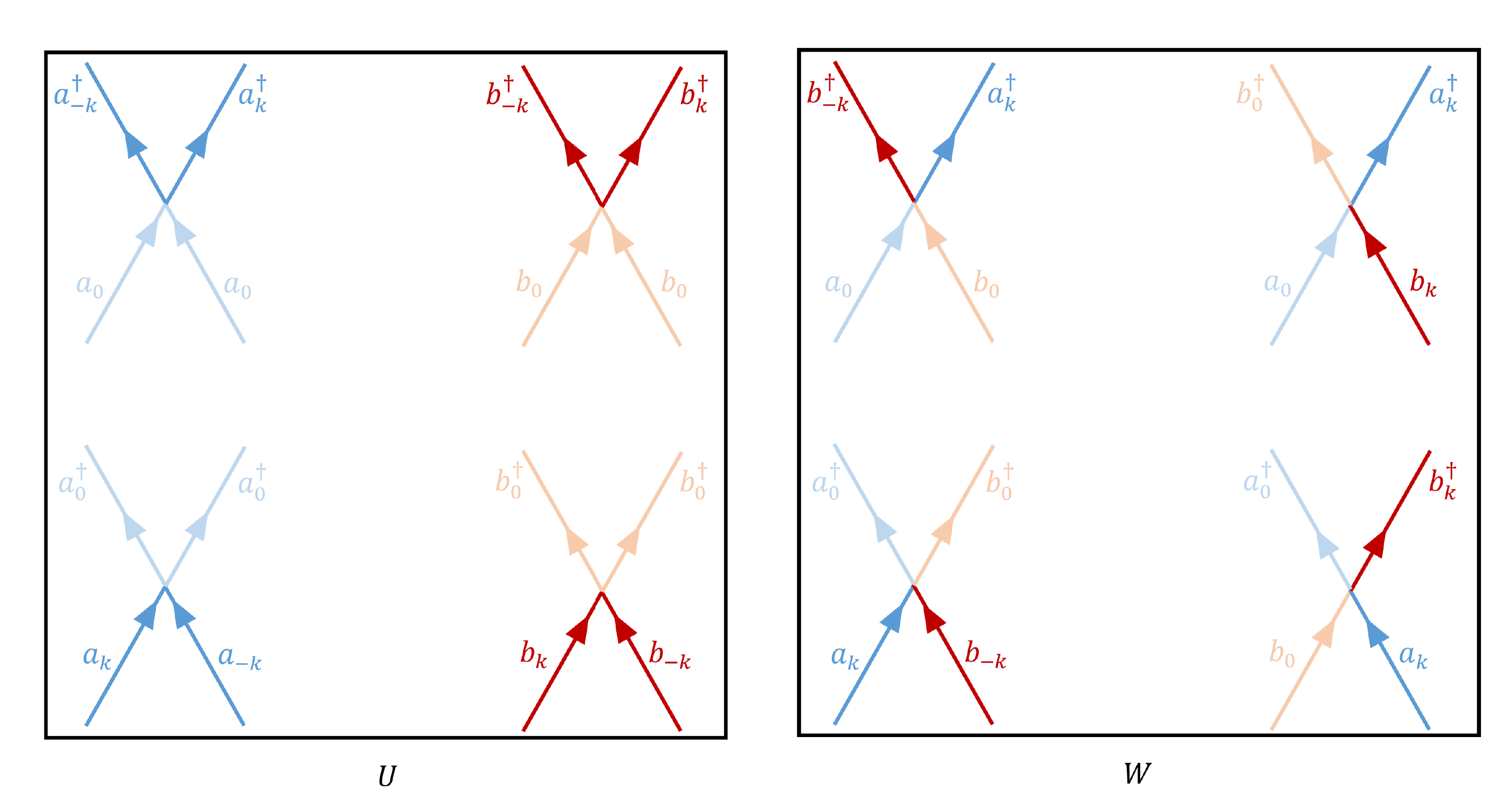}
\centering
\caption{Some important microscopic processes that occur in the mixture. Blue and red colors are used to distinguish the two species. Faded colors denote particles that belong to the two macroscopic modes, $a_0$ and $b_0$, that have been made semiclassic. Left panel: $U$ determines the creation and annihilation of excited boson-pairs within the same species. Right panel: $W$ determines the creation and annihilation of twisted boson-pairs (left side) and is responsible for scattering processes between particles of different species (right side).}
\label{fig:Feynman}
\end{figure}

The knowledge of the dynamical algebra of the Hamiltonian allows one to represent it as a linear combination  $\hat{H} = \sum_i h_i \hat{e}_i$ of algebra generators $\hat{e}_k$. For our model these coincide with the two-mode operators of the set $\cal A$. Thanks to the characteristic commutators of the algebra, the Heisenberg equation of any $\hat{e}_k$ is readily found to be
\begin{equation}
i \hbar \frac{\mathrm{d}}{\mathrm{d}t} \hat{e}_k = [ \hat{e}_k , \hat{H} ] = i \sum_m \rho_{km} \hat{e}_m , 
\label{HE}
\end{equation}
Then, the time evolution of any observable of the form
$\mathcal{O} = \sum_k o_k \hat{e}_k$ is easily determined within the algebra.
For example, from eq. (\ref{HE}) one easily discovers that operator $C_k =n_{k}-n_{-k}+ m_{k}-m_{-k}$ (the angular momentum of bosons populating the modes $\pm k$) is constant of motion. Another example concerns the number of bosons $N_*=  {A}_3+{B}_3= n_k+n_{-k}+m_k+m_{-k}$ with angular momentum 
(proportional to) $\pm k$, whose equation is
$$
i \hbar \frac{\mathrm{d}N_*}{\mathrm{d}t} 
=- 2u\left(A_+-A_- + B_+-B_- \right)-{2w}\left( K_+-K_-\right)
$$
$$
=-{2u}\left(a_{k}^\dagger a_{-k}^\dagger -a_k a_{-k} 
+  b_{k}^\dagger b_{-k}^\dagger -b_k b_{-k}\right)
$$
\begin{equation}
    \label{eq:N_*}
-{2w}\left(a_k^\dagger b_{-k}^\dagger 
+ a_{-k}^\dagger b_{k}^\dagger - a_k b_{-k} - a_{-k} b_k \right).
\end{equation}
This equation shows how the time evolution of $N_*$ depends on and implicitly defines two generalized currents of boson-pairs whose distinctive trait is to flow from the macroscopic modes $a_0$ and $b_0$ \cite{Nota_correnti} to the excited modes $a_{\pm k}$ and $b_{\pm k}$. Remarkably, while the intraspecies repulsive interaction $u$ accounts for the creation of opposite-momentum boson pairs within the \textit{same} atomic species through the terms $A_\pm$ and $B_\pm$, the interspecies coupling $w$ causes the formation of opposite-momentum boson pairs in a twisted fashion (i.e. one boson in an atomic species and one in the other) through the terms $K_\pm$. We remark that this second process is due to operators $K\pm$ which were absent in the two-ring model \cite{Noi}. Some of these microscopic processes are illustrated in Fig. \ref{fig:Feynman}. 

Another application of formula (\ref{HE}) consists in computing the dynamical equations of finer-grained observables such as the single-mode number operators $n_{\pm k}$ and $m_{\pm k}$ related to excited modes. To this end one must consider the enlarged dynamical algebra so(2,4) (including so(2,3)) in which $n_{\pm k}$ and $m_{\pm k}$ can be seen as algebra generators. The so(2,4) generators and further details are given in the Supplemental Material and in \cite{Noi}.
For the excited-mode number operator $n_k$, one finds
\begin{equation}
i\hbar \dot{n}_k = [u\, a^\dagger_k a^\dagger_{-k}
+ w( a_k^\dagger b_{-k}^\dagger +a_k^\dagger b_k )] - H.C.
\label{eq:eqn}
\end{equation}
A similar equation holds for operators $m_{k}$ in which operators $a_k$ and $b_k$ are exchanged. This equation clearly shows that a time variation of $n_k$ is caused both by intraspecies ($u \ne 0$) and by interspecies ($w \ne 0$) interaction processes. Such processes are depicted in Fig. \ref{fig:Feynman}.

\section{Dynamics on a ring trimer}
\label{sec:Mixture_on_a_trimer}

The theory and the formulas we have discussed so far are valid for 
a ring having an arbitrary site number $L$. In the present section, we investigate the simple model with $L=3$, the so-called ring trimer. This particular geometry has been thoroughly analyzed in the last decade, because of the rich scenario of nonlinear phenomena triggered by the non-integrable character of a system featuring three spatial modes \cite{Buonsante}-\cite{JJ1}. In the trimer case $k$ takes just three values, namely, $\pm 1$, and $0$ (whose relevant momentum mode will be made semiclassical in the Bogoliubov picture). Therefore, the only
sub-Hamiltonian is $\hat{H}_k=\hat{H}_1$ (see (\ref{eq:H_k})), 
associated to parameter 
$$
 \gamma_1 = 2T\left[1-\cos \left({2\pi}/{3} \right)\right] +u.
$$
To diagonalize $\hat{H}_1$, the proper choice of generalized angles is
$\mathrm{th}\theta_a =(u+w)(\gamma_1 +w)$,
$\mathrm{th}\theta_b =(u-w)/(\gamma_1 -w)$ 
and $\varphi= {\pi}/{2}$ which gives
$\hat{\mathcal{H}}=    E_0 +2 \hbar \omega \; A_3 + 2 \hbar \Omega \;B_3$
with the two characteristic frequencies 
\begin{equation}
    \label{eq:Char_freq_1}
     \omega = {\sqrt{3T(3T+2u+2w)}}/{\hbar},
\end{equation}
\begin{equation}
      \label{eq:Char_freq_2}
      \Omega ={\sqrt{3T(3T+2u-2w)}}/{\hbar} .
\end{equation}
The energy eigenvalues then read
\begin{equation}
\label{eq:Diagonal_Hamiltonian_trimer}
E(n_{\pm 1},m_{\pm 1})
=\! E_g +\hbar \omega (n_1 +n_{-1}) + \hbar \Omega (m_1 + m_{-1})
\label{EE}
\end{equation}
where $E_g = E_0 + \hbar( \omega +\Omega)$.
The good agreement of eigenvalues (\ref{EE}) with the spectrum calculated numerically is illustrated in the Supplemental Material.

The frequencies (\ref{eq:Char_freq_1}) and (\ref{eq:Char_freq_2}), and the expressions of angles $\theta_a$ and $\theta_b$ are correctly defined only in a portion of the three-dimensional parameter space $(T,U, W)$. This region of the parameter space, where the spectrum is discrete, classically corresponds to the region where the system exhibits a stable dynamics. At the border of such stability region, the energy cost to create excitations tends to zero and the system manifest unstable behaviors. From the point of view of the dynamics, in fact, one can observe the divergence of many physical observables. This aspect will be resumed in Section \ref{sec:Stability}.

\subsection{Time evolution of excitations}
In the same spirit of reference \cite{Noi}, it is possible to describe the dynamics of the momentum mode operators in terms of their expectation values. The latter are represented by the four (complex) order parameters
$$
   a_{\pm 1}=\sqrt{n_{\pm 1} }e^{i\phi_{\pm 1} }, 
   \qquad
       b_{\pm 1}=\sqrt{m_{\pm 1}}e^{i\psi_{\pm 1}}.
$$
\begin{figure}[h!]
\includegraphics[width=1\columnwidth]{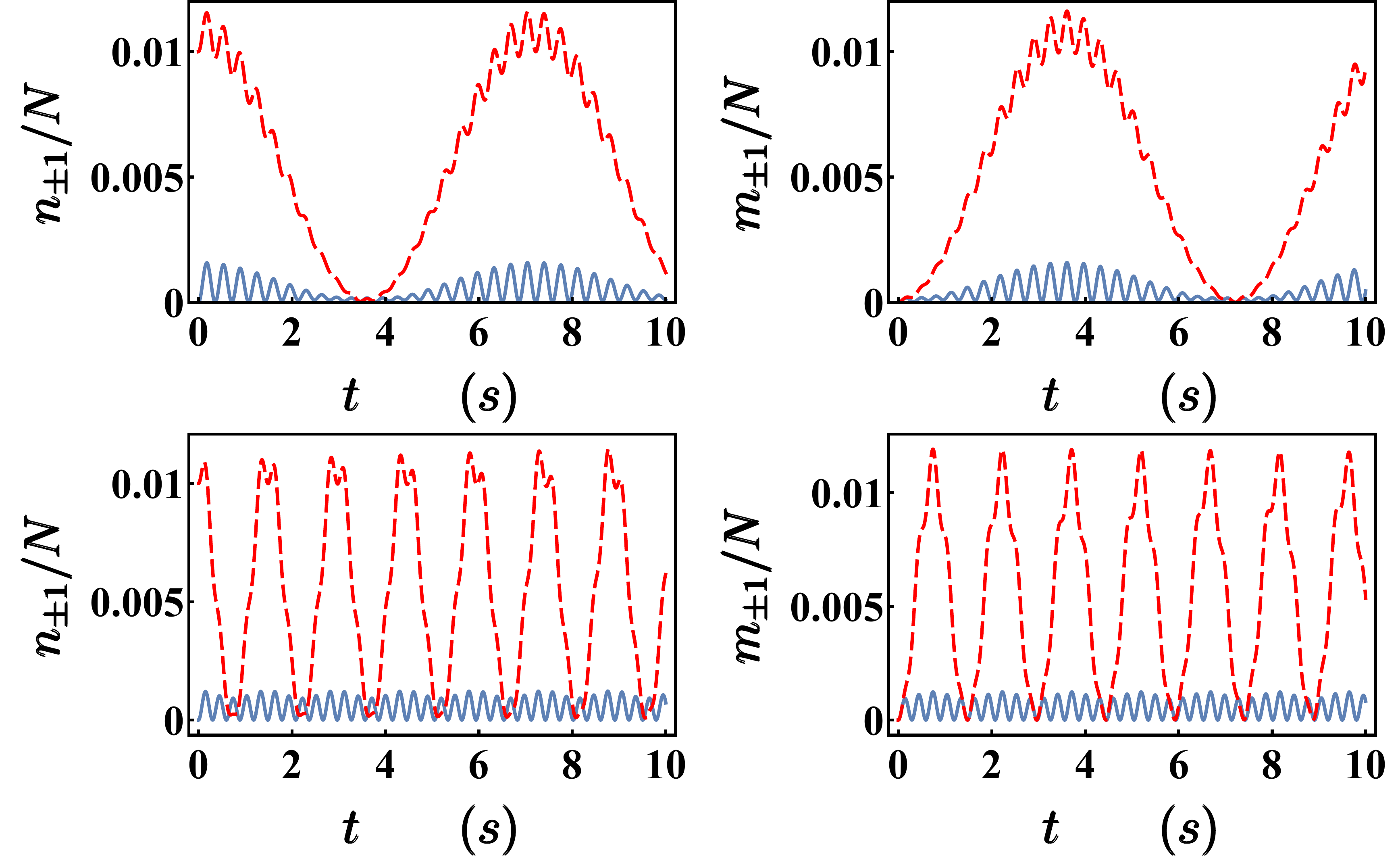}
\centering
\caption{Dynamics of excitations for $n_1(0)=10$, $n_{-1}(0)=m_{\pm 1}(0)=0$ and 
$W/U=0.19$ (upper panels) and $W/U=0.9$ (lower panels), with  $T=2$, $U=0.01$, $N=1000$ 
and $\hbar=1$. Left (right) panel concerns atomic species A (B).
Red dashed (blue) color corresponds to bosons with $k=+1$ ($k=-1$). 
The fact that $n_1(t)$ features a maximum when $m_1(t)$ 
is zero and vice-versa entails the periodic and complete AM transfer between 
the two species. Increasing $W$ strongly changes the time scale of population oscillations.
}
\label{fig:Caso_2}
\end{figure} 
Over the remainder of this subsection, we will show that one can trigger different dynamical regimes by performing different choices both for the initial state (i.e. the complex order parameter at $t=0$), and for the Hamiltonian parameters $T$, $u$, and $w$. The plots that we present below correspond to the analytic solutions of the dynamical equations (\ref{eq:eqn}) for the excited populations $n_{\pm 1}(t)$ and $m_{\pm 1}(t)$ for four different choices of initial conditions. Note that the excitation of angular-momentum modes corresponds to excite weakly-populated vortices.

{\it (1) Absence of excitations at $t=0$.} Although, at the beginning of the dynamics, none of the condensates feature any excitation, the presence of 
interactions $w \ne 0$ and $u \ne 0$ causes the periodic formation of excited bosons within the mixture, according to the formula 
$$
n_{\pm 1}(t) =m_{\pm 1} (t)= 
$$
$$
\frac{1}{2\hbar^2 }\left[    \frac{(u-w)^2}{\Omega^2}\sin^2(\Omega t) +     
\frac{(u+w)^2}{\omega^2}\sin^2(\omega t)  \right] .
$$
In other words, the fact that interactions $u$ and $w$ are non zero causes fluctuations in the vacuum
state $|n_1, \,n_{-1},\, m_1,\, m_{-1} \rangle =|0,\,0,\,0,\,0\rangle$. 
\begin{figure}[h!]
\includegraphics[angle=270,width=1\columnwidth ]{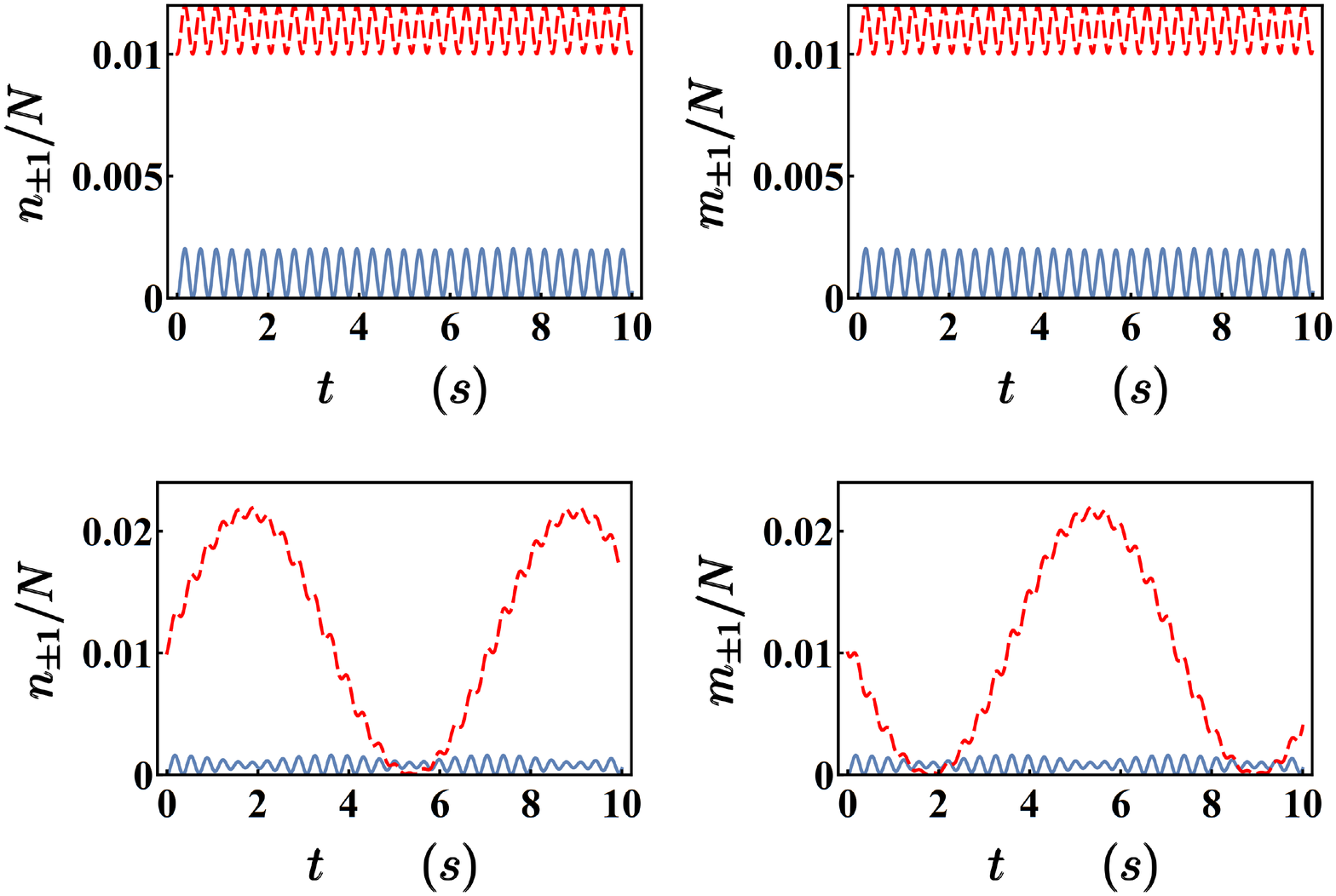}
\centering
\caption{Dynamics of excitations for $n_1(0)=m_1(0)=10$, $n_{- 1}(0)=m_{-1}(0)=0$,
and $T=2$, $U=0.01$, $W/U=0.19$, $N=1000$ and $\hbar=1$.
Left (right) panels concern species A (B). Red dashed (blue) color is used to depict 
$n_1(t)$ and $m_1(t)$ ($n_{-1}(t)$ and $m_{-1}(t)$). 
Upper panels: for an initial phase difference $\phi_1(0)-\psi_1(0)=0$ 
the system features a trivial dynamics, entailing that the inter-species  
AM transfer is suppressed.
Lower panels: Conversely, the periodic AM transfer is active
if $\phi_1(0)-\psi_1(0)$ is non-zero (e.g. equal to $\pi/2$).
}
\label{fig:Caso_4}
\end{figure}

{\it (2) Vortex-like excitation in condensate A, no excitations in condensate B.} If, at time $t=0$, one of the two atomic species exhibits a non-zero population (e.g. $n_1\neq 0$, namely a vortex excitation with $k=+1$), one observes, apart form minor quantum fluctuations (the high-frequency ripple in the figure), a periodic transfer of angular momentum (AM) between the two atomic species. This effect is illustrated in Fig. \ref{fig:Caso_2} for $W << U$ (upper panels) and $W \simeq U$ (lower panels). Of course, this case is equivalent to the case with condensate B excited and condensate A unexcited at $t=0$. 

The relevance of the inter-species parameter on the population dynamics is highlighted in the lower panels of Fig. \ref{fig:Caso_2}: Increasing $W$ clearly shows how the periodic collapse and revival of populations $n_{-1}$, $m_{-1}$ is replaced by an essentially regular oscillation while the oscillation of $n_{+1}$, $m_{+1}$ as well as the AM transfer take place on a much smaller time scale.
The same periodic AM transfer is also observed when, in addition to a vortex, 
the initial configuration includes an anti-vortex in the same species.

\begin{figure}[h!]
\includegraphics[width=1\columnwidth]{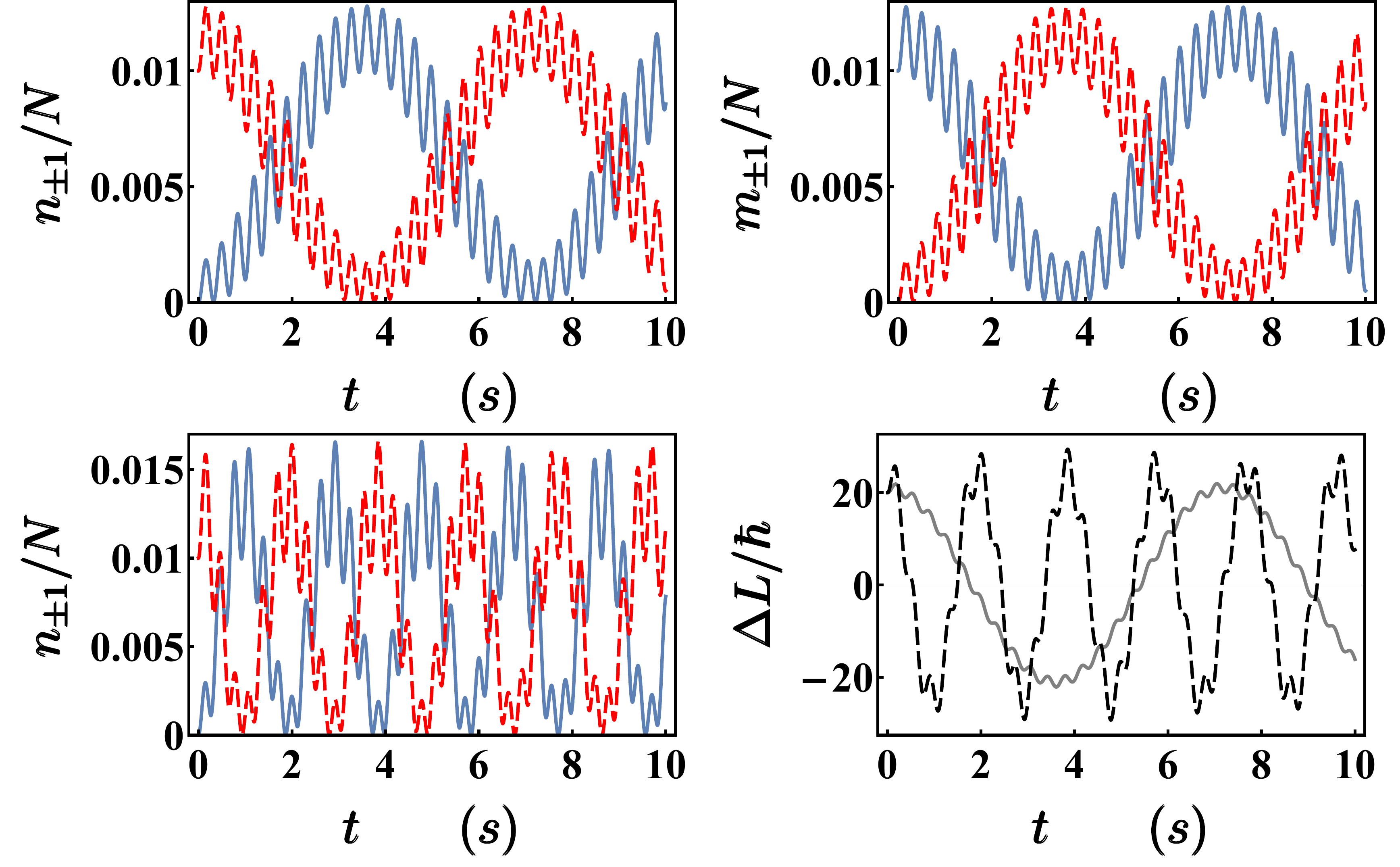}
\centering
\caption{Upper panels: Dynamics of excitations for $n_1(0)=m_{-1}(0)=10$, $n_{-1}(0)=m_{1}(0)=0$
and $W/U= 0.19$, $T=2$, $U=0.01$, $N=1000$ and $\hbar=1$. The left (right) panel concerns species 
A (B). Red dashed (blue) color corresponds to bosons with $k=+1$ ($k=-1$). Note, up 
to quantum fluctuations (high frequency ripple), the periodic and complete transfer of excitations
between the two species.
Lower left panel: for $W/U= 0.9$ the frequency of oscillations increases (the color code is the same
as in the upper panels). Right lower panel: Gray (dashed black) lines describes the time evolution of
the inter-species AM difference for $W/U= 0.19$ ($W/U= 0.9$).
}
\label{fig:Caso_5}
\end{figure} 
{\it (3) Vortex in condensate A and equal vortex in condensate B.} Let us assume that, at $t=0$, species A exhibits a weakly-populated vortex while an equal vortex is present in species B. This is an interesting situation, because different initial phase differences $\phi_1(0)-\psi_1(0)$ can trigger different dynamical regimes. Namely, if the complex quantities $a_1$ and $b_1$ are \textit{in phase} at $t=0$, the two condensates seem to be decoupled and just feature quantum fluctuations (upper panels of Fig. \ref{fig:Caso_4}). Conversely, an initial non-zero phase difference makes it possible a periodic AM exchange between the two condensates (lower panels of Fig. \ref{fig:Caso_4}). Once more, we note that the same dynamics is observed by exchanging A and B at $t=0$.

{\it (4) Vortex excitation in condensate A, and equal anti-vortex in condensate B.} Finally, let us consider the case where a weakly-populated vortex of species A ($n_1\neq 0$) is superimposed to an equal but counter-propagating anti-vortex of species B ($m_{-1}\neq 0$). It turns out that the two kinds of excitations are periodically transferred from a species to the other (see upper panels of Fig. \ref{fig:Caso_5}). 
The left lower panel of Fig. \ref{fig:Caso_5} confirms the relevance of parameter $W$: As discussed in the case ({\it 2}), increasing $W$ implies that the population oscillations and the AM transfer take place on a much smaller
time scale. In particular, the right lower panel of Fig. \ref{fig:Caso_5} illustrates the AM transfer through the quantity $\Delta L= n_{+1}- n_{-1} -(m_{+1}- m_{-1})$ where the signs $\pm$ in front of the mode population takes into account the clockwise (or anti-clockwise) rotations of bosons. 

It is important to note that the equation governing the dynamics of the AM transfer $\Delta L$ in the cases $(\it 2)$-$(\it 4)$
$$
i \hbar \frac{d \Delta L}{dt} =
2w ( a_{k}^\dagger b_{-k}^\dagger + a_{-k} b_{k}
+ a_{k}^\dagger b_{k} + a_{-k} b_{-k}^\dagger - H.C. ) 
$$
clearly shows how the microscopic processes involved by operators $S_\pm$ and $K_\pm$ both contribute to activate the AM transfer. This circumstance is due to the fact that, unlike the two-ring ladder model, where the transfer is controlled by the inter-ring tunneling, in the current model the transfer is triggered by the inter-species interaction.

\section{Towards dynamical instability}
\label{sec:Stability}
\begin{figure}[h!]
\includegraphics[width=1\columnwidth]{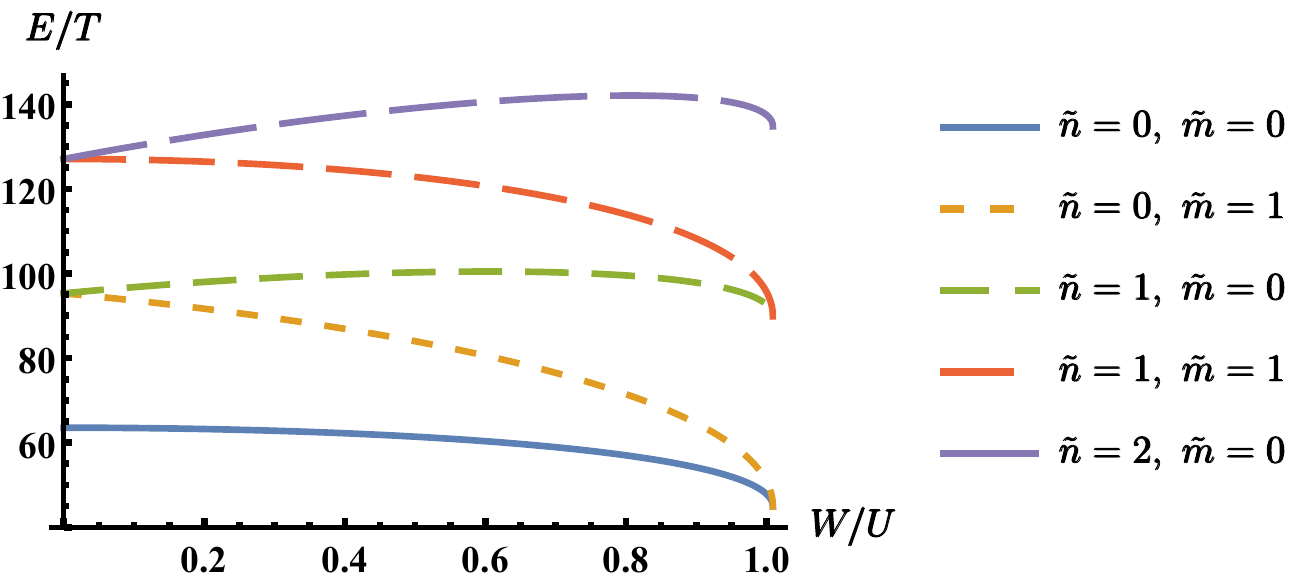}
\centering
\caption{Energy levels vs interspecies interaction $W$ for a trimer. If the two atomic species do not interact ($W=0$), the two characteristic frequencies $\omega_1$ and $\Omega_1$ coincide. When $W/U$ approaches the limiting value $1.009$, one can observe the collapse of the energy spectrum relevant to the second generalized harmonic oscillator. Parameters $T =2$, $U=1$, $N=1000$ have been chosen.}
\label{fig:Spectral_collapse}
\end{figure}

In this section we resume the discussion on the diagonalization procedure of Section \ref{sec:Dynamical_algebra} and comment on the stability of the system. One can observe that the angles associated to generalized rotations, $\theta_a$ and $\theta_b$, (see Equations (\ref{eq:Rotation_angles})) are subject to some \textit{constraints}. The same constrains can also be found in relation to the diagonal Hamiltonian (\ref{eq:Diagonal_Hamiltonian}), i.e., in the expressions of characteristic frequencies (\ref{eq:omega_k}) and (\ref{eq:Omega_k}). Recalling that $T$, $U$ and $W$ have been assumed to be non-negative, all the dynamical results and the solution scheme that we have discussed up to now are valid provided that
\begin{equation} 
w<T(1-c_k)+u .
\label{threshold}
\end{equation} 
When $w$ approaches this upper limiting value, the energy spectrum collapses, i.e., the energy difference between two adjacent energy levels tends to vanish (Fig. \ref{fig:Spectral_collapse} well illustrates the collapse).
As a consequence, each $\hat{H}_k$ is associated to a specific limiting value for $w$, namely $T(1-c_k)+u$. Recalling that $1-c_k=1-\cos(a\tilde{k})$ is an increasing function in the interval $k\in \left[0,({L-1})/{2}\right]$, the global limiting value is always found at the smallest momentum, i.e., for $k=1$, irrespective of site number $L$. In other words, increasing $w$, the sub-Hamiltonian which could first be affected by the spectral collapse is $\hat{H}_1$, no matter the number of ring-lattice sites $L$.
\begin{figure}[h!]
\includegraphics[width=1\columnwidth]{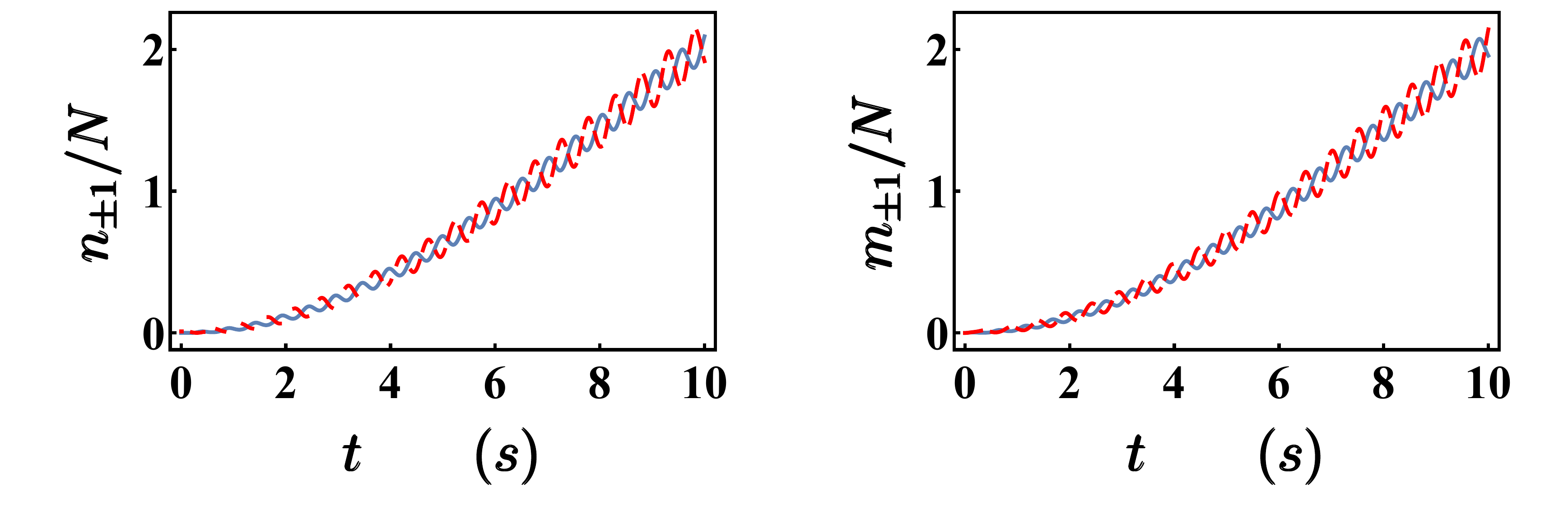}
\centering
\caption{Time evolution of populations for $U=0.01$, $T=2$, $W/U=1.895$, $\hbar=1$, $N=1000$, $n_1(0)=10$ and  $n_{-1}(0)=m_{\pm 1}(0)=0$. Notice that $W$ is close to the critical value $W_c$, i.e. $W/W_c=0.997$. $n_{1}(t)$ and $m_1(t)$ are depicted in red, while $n_{-1}(t)$ and $m_{-1}(t)$ are represented in blue. All the excited boson populations quickly grow and soon get unphysical.}
\label{fig:Divergence}
\end{figure}

With reference to the most elementary closed circuit, the trimer, the stability condition becomes
\begin{equation}
w<\frac{3}{2}T + u,
\label{trim}
\end{equation}
(namely, $W <W_c:= 9T/(2N) +U$ in terms of the model parameters) which guarantees the correctness of formulas (\ref{eq:Char_freq_1}), (\ref{eq:Char_freq_2})
and (\ref{eq:Diagonal_Hamiltonian_trimer}). As far as excited populations, they reveal 
a \textit{diverging} evolution when one approaches the border of the stability region. 
Fig. \ref{fig:Divergence} clearly depicts this situation. 
In view of this, the inequality (\ref{trim}) is particularly significant in that, 
in addition to setting the limits of validity for our model,
it can be seen as the hallmark of a dynamical phase transition \cite{Jack}, \cite{VP}.
Finally, we note how the more general inequality (\ref{threshold}) reproduces, for
a generic ring lattice, the {\it spectral-collapse conditions} related to the demixing effect for a mixture in a 2-well potential \cite{spect}.

\section{Concluding remarks}
\label{sec:Concluding_remarks}
We have studied the weak excitations of a two-species bosonic mixture confined in a BH ring. 
In Section \ref{sec:Model_presentation} we proved that, after enacting the well-known Bogoliubov procedure, the model Hamiltonian reduces to the sum of $(L-1)/2$ independent sub-Hamiltonians $\hat{H}_k$ related to \textit{pairs} of (opposite) momenta. In particular, we showed that each $\hat{H}_k$ belongs to a specific dynamical algebra, the algebra so(2,3).
Thanks to the knowledge of the dynamical algebra, in Section \ref{sec:Dynamical_algebra}, we diagonalized the effective Hamiltonian, determined the energy spectrum, computed the Heisenberg equations for various physical observables, and highlighted microscopic processes characterizing the mixture. 

In Section \ref{sec:Mixture_on_a_trimer}, we applied the proposed solution scheme to the simple but non trivial three-well ring, the BH trimer, featuring $r=0$ as macroscopically occupied momentum mode. 
The corresponding energy spectrum was determined and shown to provide the two frequencies that characterize the trimer dynamics.
We computed the time evolution of excited populations for different choices of the model parameters and different values of initial conditions. More specifically, we pointed out 
the presence of fluctuations in the vacuum-state, 
the possible coherent periodic transfer of angular momentum between the two species 
and its relation with their initial \textit{phase difference}. 
Also, we emphasized the influence of the interspecies interaction $W$ on the population dynamics by comparing the population oscillations in the case when $W$ is smaller than $U$ and when $W \simeq U$. We showed that increasing $W$ makes the AM transfer faster.

As a conclusion, in Section \ref{sec:Stability}, we identified the region where the system is stable, and observed that, for $w\to u+{3}T/{2}$, the system approaches instability, a possible signature of the mixing-demixing phase transition \cite{spect}. This issue, the analysis of attractive interactions and of strong interspecies repulsions, will be considered in a future work.  
%
%


\begin{thebibliography}{40}
\bibitem{Nota_correnti} One should remember that, in the Bogoliubov scheme,  $a_0$ and $b_0$ have been incorporated in the effective interaction strengths $u$, $w$ in terms of the parameters $\sqrt N$ and 
$\sqrt M$.
\end{thebibliography}
\end{document}